\documentstyle[11pt,aaspp4]{article}	

\slugcomment{Submitted to the {\it P. A. S. P.}}

\lefthead{Lauer}
\righthead{Undersampled Point Spread Functions}

\def\shah{\rm III}
\begin{document}
\title{The Photometry of Undersampled Point Spread Functions}

\author{Tod R. Lauer}
\affil{National Optical Astronomy
Observatories\altaffilmark{1}, P.~O. Box 26732, Tucson, AZ 85726}
\affil{Electronic mail: lauer@noao.edu}

\altaffiltext{1}{The National Optical Astronomy Observatories are
operated by the Association of Universities for Research in Astronomy, Inc.,
under cooperative agreement with the National Science Foundation.}

\begin{abstract}

An undersampled point spread function may interact with the microstructure
of a solid-state detector such that the total flux detected can depend
sensitively on where the PSF center falls within a pixel.
Such {\it intra-pixel} sensitivity variations will not be corrected
by flat field calibration and may limit the accuracy
of stellar photometry conducted with undersampled images,
as are typical for {\it Hubble Space Telescope} observations.
The total flux in a stellar image can vary by up to 0.03 mag in F555W WFC
images depending on how it is sampled, for example.  For NIC3, these
variations are especially strong, up to 0.39 mag, strongly limiting its use
for stellar photometry.
Intra-pixel sensitivity variations can be corrected for,
however, by constructing a well-sampled PSF from a dithered data set.
The reconstructed PSF is the convolution of the
optical PSF with the pixel response.  It can be evaluated at any desired
fractional pixel location to generate a table of photometric corrections
as a function of relative PSF centroid.
A caveat is that the centroid of an undersampled PSF can also
be affected by the pixel response function, thus sophisticated centroiding
methods, such as cross-correlating the observed PSF with its fully-sampled
counterpart, are required to derive the proper photometric correction.

\end{abstract}


\section{Introduction}

The study of stellar populations has been revolutionized by the
techniques of crowded-field photometry applied to {\it Hubble Space
Telescope} images.  Ironically, however, nearly all {\it HST} images
are undersampled and are thus not optimal for such problems.
The information missing from undersampled images makes it difficult to detect
faint sources, eliminate cosmic ray hits, register different exposures,
accurately represent the stellar point spread function (PSF), and so on.
These problems are well known to practitioners of crowded-field photometry,
and can be partially countered by using impressive
software packages such as ALLFRAME (\markcite{all}Stetson 1994), DoPHOT
(\markcite{do}Schechter, Mateo, \& Saha 1993), or CCDCAP
(\markcite{mig}Mighell 1997),
which are specially tuned to extract information from multi-image
data sets that may be poorly represented in any single image.
The effects of undersampling on the detected flux of a stellar image, itself,
may be less apparent, but are potentially important.
While one might assume that
a photon will generate an electron that will land in one pixel or another,
regardless of how the sampling is done, in reality the complex microstructure
of a CCD or any other solid-state detector may cause its response
to vary significantly over the area of a single pixel.
In this paper I present a method to calibrate and correct
such {\it intra-pixel} sensitivity variations, with particular
application to {\it HST} WFPC2 and NIC3 images.

In a well-sampled image the pixel spacing or sampling frequency is
sufficient to completely characterize its structural content
on all spatial scales;
in a poorly sampled image fine-scale structure may be
present that can interact with the yet higher spatial
frequencies associated with the detector microstructure.
\markcite{jdo}Jorden, Deltron, \& Oates (1994) used a pinhole projector to
measure the intra-pixel response of a variety of CCDs.  Their experiments
showed that the total detected flux of an undersampled PSF can
vary strongly ($>\pm10\%$) with centering within a pixel, source color,
and differences among the gate structure of the various CCDs.
Front-illuminated devices showed the strongest effects, but
significant variations could still be seen with rear-illuminated CCDs.
Not surprisingly, the response varied differently as a function
of the row versus column position, given the anisotropic structure of CCDs.

Real astronomical cameras are difficult to calibrate with such
laboratory experiments, given the sensitivity to the degree
of PSF undersampling, but intra-pixel effects may be detected
through a ``dithered'' set of images of a star field, that is images
slightly offset from each other by a fraction of a pixel.
\markcite{holtz}Holtzman et al. (1995) used such images of
the $\omega$ Cen globular cluster to show that the detected stellar
flux varied by a few percent as a function of fractional column
position in the WFPC-2 CCDs.
While such a small effect may of little concern for most WFPC-2 programs,
the situation is far different in NIC3 images.
Here the strong undersampling, coupled with the particular microstructure
of the NICMOS arrays (which are not CCDs) causes the detected flux
in a stellar image to vary by up to $\pm 0.2$ mag in the bluest
(and hence most poorly sampled) colors.  Clearly, this strongly
limits use of NIC3 for stellar photometry work.

Calibration of the camera response to undersampled images
can be done in a variety of ways, given a dithered image set of point sources.
My approach is to reconstruct a fully sampled
``superimage'' from the data set, which can be then used to make ``observed''
images true to the original sampling, but with any desired spatial offset;
one can then simply measure how the integrated flux of a point source
varies with its fractional offset with respect to the pixel grid.
Any single image can be expressed as
\begin{equation}
I(x,y)=O(x,y)\ast P(x,y)\left(\shah(x,y) \ast {\cal R}(x,y)\right),
\end{equation}
where $O$ is the intrinsic projected appearance of the astronomical
field being imaged, $P$ is the PSF due to the telescope and camera optics,
$\shah(x,y)$ is a two-dimensional array of sampling points
\begin{equation}
\shah(ax,ay)\equiv{1\over{\vert a\vert^2}}
\sum_{i=-\infty}^{+\infty}
\sum_{j=-\infty}^{+\infty}\delta\left(x-{i\over a}\right)
\delta\left(y-{j\over a}\right),
\end{equation}
and $\ast$ means convolution.
The critical term for the present discussion is
${\cal R}(x,y),$ the generally unknown spatial response of the pixel, itself.
This term not only includes the sensitivity response as a function
location within the pixel, but also any diffusion of photons
or photoelectrons within the device --- its extent may thus be
larger than that of a single pixel.

Producing a dithered image set by stepping the detector a
fractional amount in $x$ and $y$ can be used to produce
a more finely sampled superimage.  When the dithers are done
in a regular $N\times N$ pattern of subpixel steps (of relative
size $1/N$),
\begin{eqnarray}
\label{interlace}
I_S(x,y)&=&O(x,y)\ast P(x,y)\sum_{i=0}^{N-1}\sum_{j=0}^{N-1}
\shah\left(x-{i\over N},y-{j\over N}\right) \ast {\cal R}(x,y), \nonumber \\
&=&\left(O(x,y)\ast P(x,y)\ast {\cal R}(x,y)\right)\shah(Nx,Ny).
\end{eqnarray}
The new superimage thus has an effective PSF
\begin{equation}
\label{Ps}
P^\prime(x,y)=P(x,y)\ast {\cal R}(x,y).
\end{equation}
For severely undersampled images, ${\cal R}$ may actually be more important
than the core structure of $P$
for setting the effective resolution of $P^\prime.$
If $N$ is large enough such that $P^\prime$ is fully sampled
($N=3$ for WFPC-2 is sufficient),
then $P^\prime$ can be interpolated to any desired location
with respect to the original undersampled pixel array.
Drawing every $N$th pixel in $x$ and $y$ from the superimage
generates an image as would have been observed at the given position.
Comparing the integrated flux in the interpolated-undersampled
PSF to that in $P^\prime$ thus allows the photometric effects of undersampling
to be measured for any desired fractional location with respect
to the original grid.  Note that ${\cal R}$ need not be determined itself,
since it is implicitly included in $P^\prime,$ and its effects depend
critically on the structural content of $P$ in any case.

The tricky step is generating a fully-sampled superimage.
While it may be possible to step the detector position
in a regular subpixel sampling grid, in practice this may be difficult.
Sub-pixel dithers have been used in many WFPC-2 programs,
for example, but were often
not executed with enough precision to fall on a regular pattern.
In this case, the simple interlacing of the dithered image set
implied by equation (\ref{interlace}) cannot be done.
Reconstruction of a well-sampled superimage from a set of undersampled
images is a difficult problem if the geometric relationships
among the images are complex.
For the simple case, however, where the images in the set are related
by purely translational (if arbitrary) offsets, the sampling grid is spatially
constant, and the intrinsic object, PSF, and detector properties
do not vary over the set, it is possible to construct a superimage
in closed form through a complex linear combination of the images
in Fourier space (\markcite{far}Lauer 1999).
While these requirements may sound highly and
perhaps impossibly idealized, in practice they can be realized in
{\it HST} observations with both WFPC2 and NICMOS, for example,
if the dither offsets are relatively small with respect to any changes
in the angular pixel sampling frequency, and the observations are
obtained over a short enough time span such that any focus changes,
source variability, and so on, are unimportant.
In passing, I note that the {\it Drizzle} algorithm
(\markcite{driz}Fruchter \& Hook 1999) has proven to be
an extremely popular and versatile tool for building superimages
from {\it HST} data sets; however, {\it Drizzle} does not guarantee
that the superimage is well-sampled, and since it also introduces
a variable blurring function on the finest scales, I am concerned
that it may not be suited to the present application.

\section{Calibrating the Photometry of an Undersampled PSF}

\subsection{Construction of a Well-Sampled PSF}

In greater detail, the method of PSF reconstruction advocated
here involves stacking the set of dithered images in the Fourier domain.
The Fourier transform of a discretely sampled data set is periodic,
repeating to $\pm\infty$ in both $x$ and $y.$
When an image is undersampled, or aliased, the higher order ``satellites''
in the Fourier domain overlap with and contaminate the fundamental transform.
This contamination cannot be eliminated in any single image, but
as the object being imaged is shifted with respect to the pixel grid,
the phases within the Fourier satellites vary.
With a sufficient number of dithered images, each having different
phases, the aliasing can be eliminated algebraically.
The sampling frequency in the final superimage is determined by
the spatial scale at which $P^\prime$ no longer has significant power.
This approach is nicely summarized by \markcite{fft}Bracewell (1978)
for the case of one-dimensional data;
I present a tutorial on its extension to images, with particular
application to {\it HST} WFPC-2 images, in \markcite{far}Lauer (1999).
The attractive features of this method are that the superimage
is well-sampled, there are no arbitrary parameters controlling
its construction, and there is no blurring at the Nyquist scale.
The method requires a minimum of $N^2$ images with non-degenerate
dithers to construct a superimage with $N\times N$ subsampling;
when more images are available the superimage is overdetermined
and becomes the best-fit to the dither set.

The {\it HST} WFPC-2 and NICMOS imagers provide contrasting
test cases for exploring the effects of intra-pixel sensitivity variations.
Intra-pixel effects are subtle in the WFC chips of WFPC-2,
but severe in the NIC3 camera.
\begin{figure}[hbtp]
\plotone{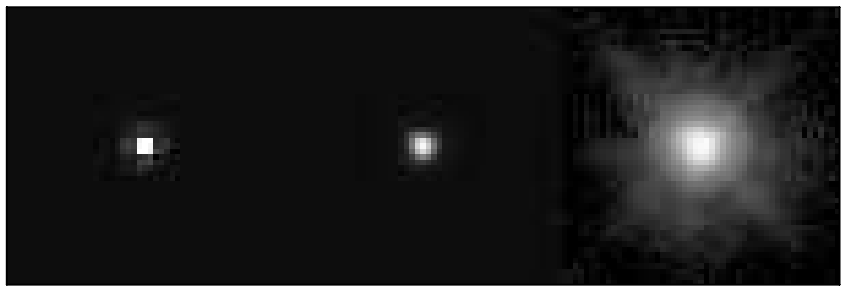}
\caption{Reconstruction of an {\it HST} WFC PSF with $3\times3$
subsampling is shown based on 20 dithered F555W images of a
star in $\omega$ Cen.  The image at left shows a linear stretch of
the PSF with the original sampling ($0\farcs10$ pixels).
The central image shows the reconstructed PSF with the same intensity stretch
with the full $3\times3$ subsampling.
The last image is a logarithmic stretch (with dynamic range 3.5 in log units)
of the reconstructed PSF. Each subimage is $1\farcs7\times1\farcs7$ in size.}
\label{fig:wfc_psf_V}
\end{figure}
\begin{figure}[hbtp]
\plotone{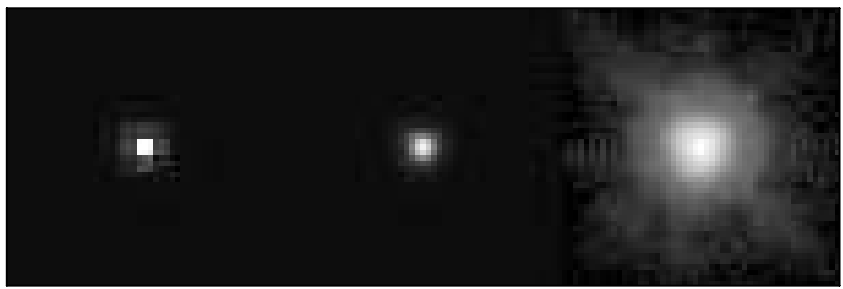}
\caption{Reconstruction of an {\it HST} WFC PSF in the F814W
filter, as in Figure 1.  Both figures show the same star.}
\label{fig:wfc_psf_I}
\end{figure}
\begin{figure}[hbtp]
\plotone{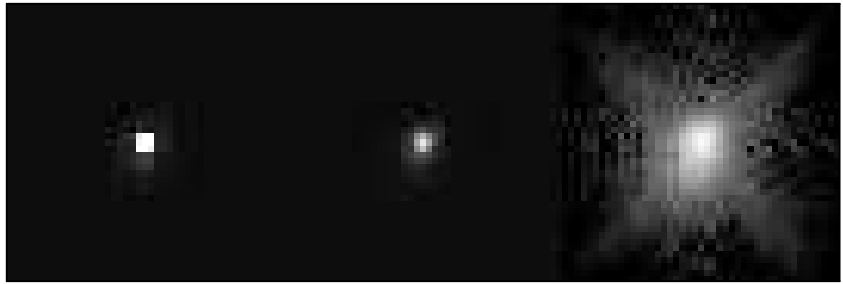}
\caption{Reconstruction of an {\it HST} NIC3 F110W (J-band) PSF with $3\times3$
subsampling based on dithered images obtained as part
of the {\it Hubble Deep Field South} program.  The star shown is the brightest
source in the NIC3 HDFS field.
The image at left shows a linear stretch of
the PSF with the original sampling ($0\farcs20$ pixels).
The central image shows the reconstructed PSF with the same intensity stretch
with the full $3\times3$ subsampling.
The last image is a logarithmic stretch (with dynamic range 3.5 in log units)
of the reconstructed PSF. Each subimage is $3\farcs0\times3\farcs0$ in size.}
\label{fig:nic_psf_J}
\end{figure}
\begin{figure}[hbtp]
\plotone{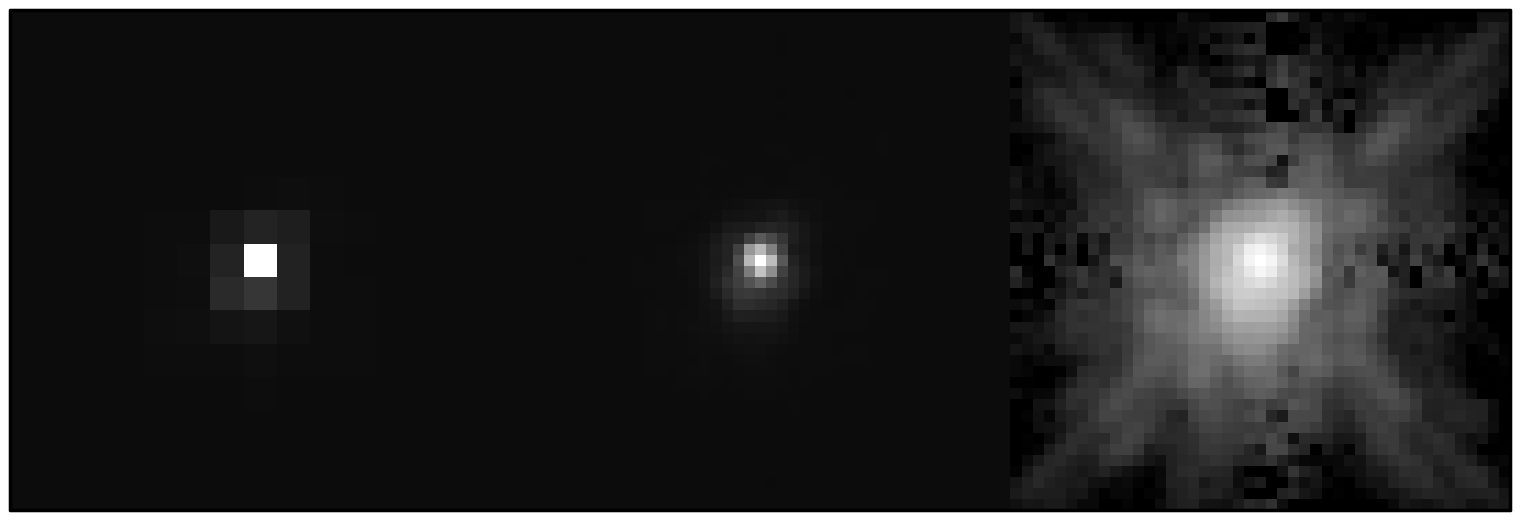}
\caption{Reconstruction of an {\it HST} NIC3 PSF in the F160W (H-band)
filter, as in Figure 3.  Both figures show the same star.}
\label{fig:nic_psf_H}
\end{figure}
Figures \ref{fig:wfc_psf_V} to \ref{fig:nic_psf_H} show PSFs reconstructed
with $3\times3$ subsampling
for the V (F555W) and I (F814W) filters in the WFC camera of WFPC-2, and
the J (F110W) and H (F160W) filters for NIC3.
The WFPC-2 PSFs were constructed from dithered observations of $\omega$ Cen
(STScI program 4819) actually obtained by the WFPC-2 IDT for the purpose of
understanding variations in the WFPC-2 PSF as a function of pixel location.
The image set consists of 20 images in each of
two filters, F555W, and F814W.
The dither pattern consists of $0\farcs025$ steps in the row and
column directions, the 20 images mapping out a
$0\farcs075\times0\farcs1$ rectangle with a square grid.

The NIC3 PSFs were constructed from images obtained for the
{\it Hubble Deep Field South} program.
The image set comprised 146 exposures, of which 98 were used
(the star in the discarded images was either out of the field,
or too close to its border).
The PSFs shown are from the brightest star in the HDF-S field.
The HDF-S dither pattern consisted both of small and large angular offsets,
largely dictated by the needs of the other cameras used in the HDF-S program.
While this data set was fine for the present purposes, in many
ways the dither pattern was far from optimal, an issue that I
will discuss in further detail below.

\subsection{Mapping the Photometric Variation of an Undersampled PSF}

It is simple to calibrate the effects of intra-pixel sensitivity
variations and undersampling on a PSF, once $P^\prime$ has been constructed.
Since $P^\prime$ is well-sampled, its centroid can be shifted to any
desired fractional pixel location, without loss of resolution or information.
Once shifted, coarse samples can be drawn from $P^\prime$ to simulate
a PSF, $P_0(\delta x,\delta y),$ as would be observed at that location.
$P_0$ is thus
\begin{equation}
\label{PO}
P_0(\delta x,\delta y)=P^\prime(x-\delta x,y-\delta y)\shah(x,y),
\end{equation}
where the $\shah$ function refers to the spacing of the detector
(rather than subsampled) pixels.
The photometric error, $\epsilon(\delta x,\delta y),$
at the given offset is
\begin{equation}
\label{error}
\epsilon(\delta x,\delta y)=\int_{-\infty}^{+\infty}\int_{-\infty}^{+\infty}
P_0(\delta x,\delta y)dx~dy\Biggl/
\int_{-\infty}^{+\infty}\int_{-\infty}^{+\infty} P^\prime(x,y)dx~dy\ -\ 1
\end{equation}
One can thus systematically map $\epsilon$ over the entire domain
of fractional centroid offsets at any desired resolution;
the map essentially consists of a lookup table of photometric
offsets to be applied to a reduced photometric data set.
It is critical to use a method of interpolation
for $P^\prime$ that does not degrade the resolution;
I do this with sinc-function interpolation, which
is the theoretically appropriate sampling kernel for well-sampled data.
Lastly, I emphasize that no ``integration over a pixel'' is
included in equation (\ref{PO}), nor should be.
Remember, $P^\prime$ already reflects convolution of the optical PSF
with the detector pixel response, thus this integration has already
implicitly taken place.

\begin{figure}[thbp]
\plotone{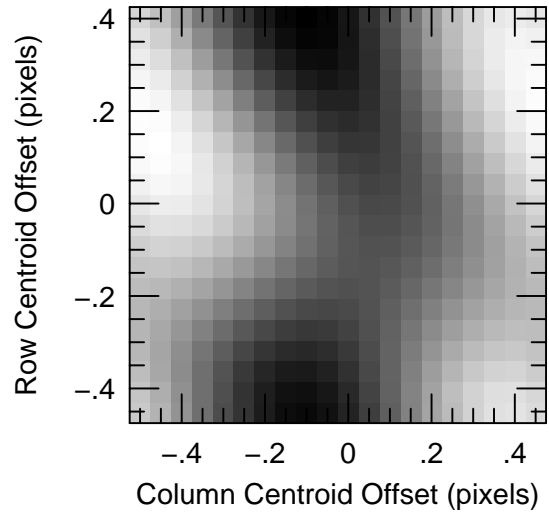}
\caption{The photometric error caused by undersampling
is shown for the V-band (F555W) WFC PSF (presented in Figure 1)
as a function of fractional pixel location of the PSF centroid.
The area shown is that of a single pixel, corresponding
to centroid offsets of $-1/2<\delta x<1/2,$ $-1/2<\delta y<1/2,$
in units of the original WFC pixel; the position of no PSF offset,
that is a PSF centered precisely on a WFC pixel is at the center of the map.
Results are presented in steps of 0.05 pixels in $x$ and $y$.
The gray scale is linear with the stretch set to the full range of
photometric error measured, with white corresponding to 0.016 mag
of excess flux, and black to a 0.014 mag deficit.  The
maximal flux is actually detected at the $x$ margins, corresponding to a PSF
centered between columns on the WFC CCDs.}
\label{fig:v_error}
\end{figure}

\begin{figure}[bhtp]
\plotone{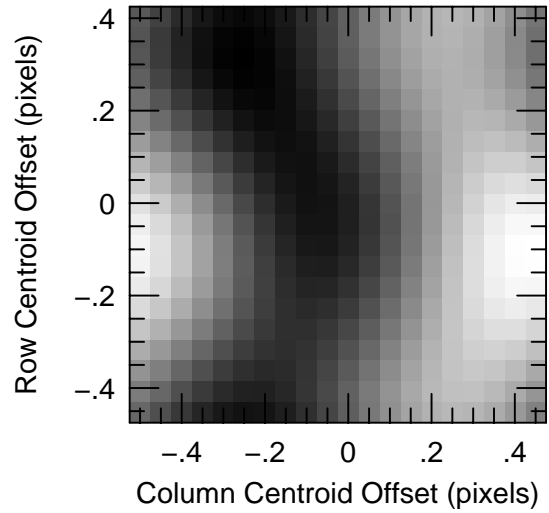}
\caption{The photometric error caused by undersampling
is shown for the I-band (F814W) WFC PSF (presented in Figure 2).
The stretch now corresponds to 0.013 mag excess to 0.011 mag deficit.}
\label{fig:i_error}
\end{figure}

\begin{figure}[htbp]
\plotone{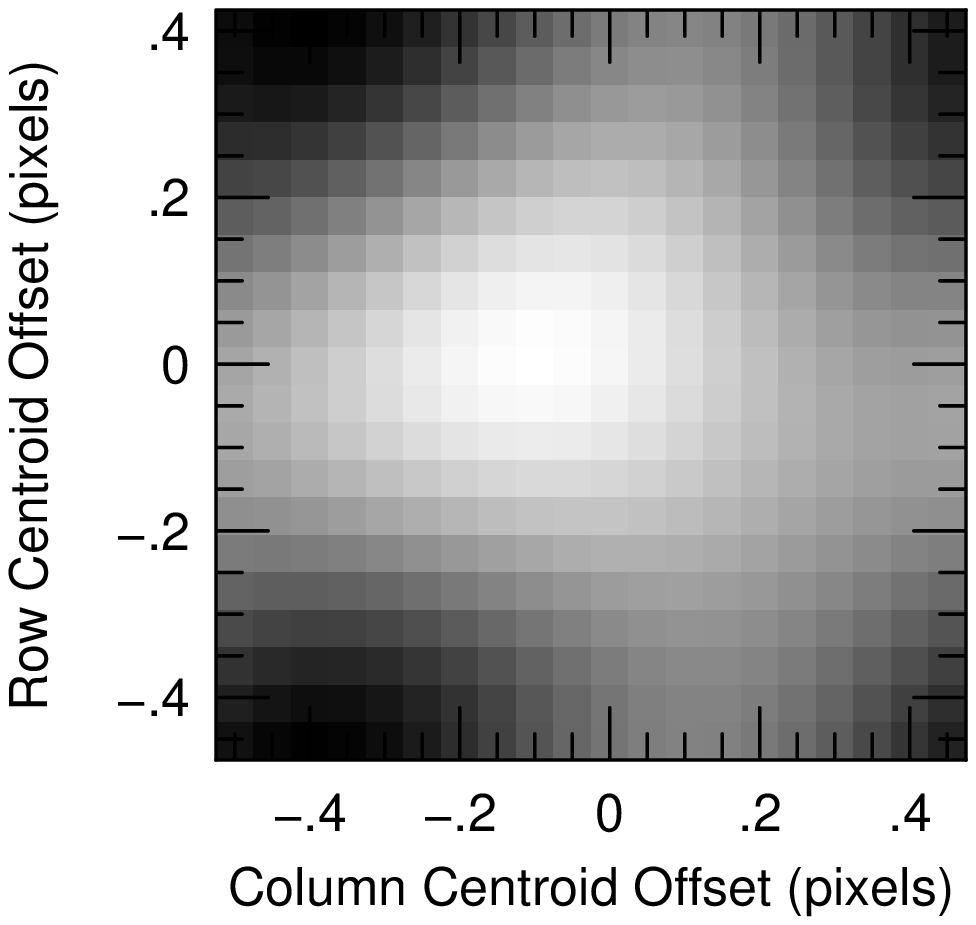}
\caption{The photometric error caused by undersampling
is shown for the J-band (F110W) NIC3 PSF (presented in Figure 3).
The stretch now corresponds to 0.22 mag excess to 0.17 mag deficit.}
\label{fig:j_error}
\end{figure}

\begin{figure}[htbp]
\plotone{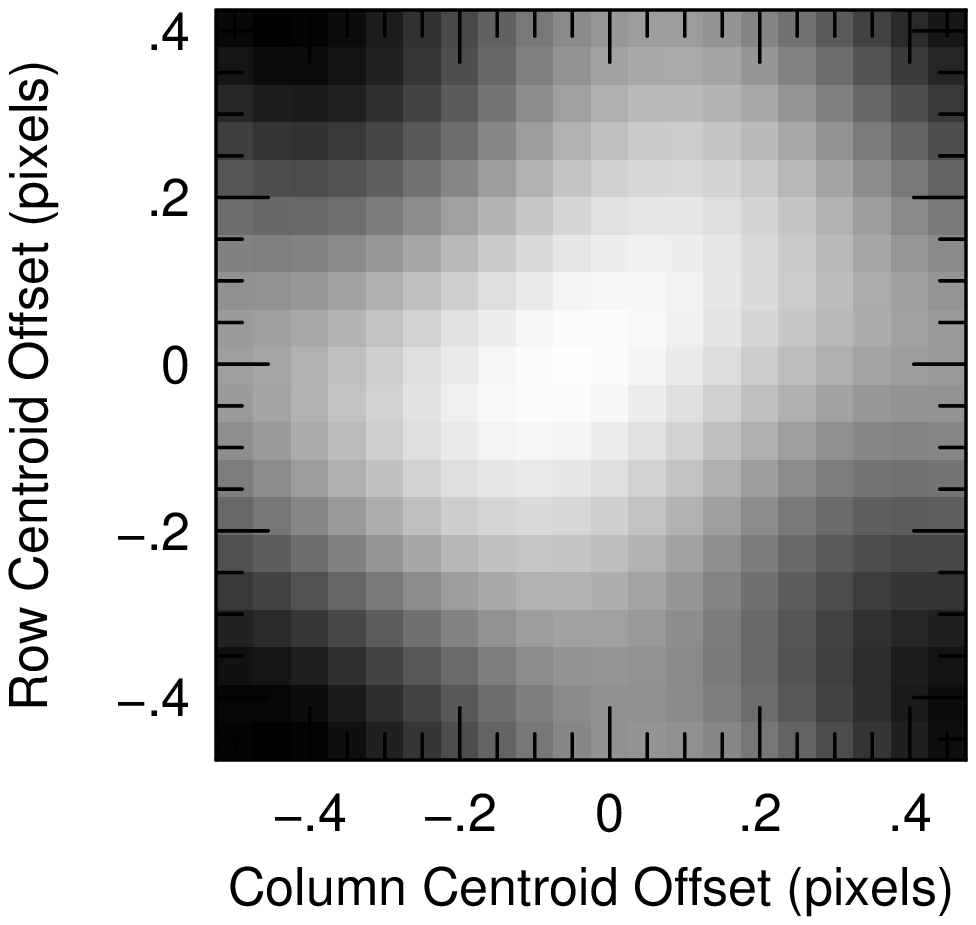}
\caption{The photometric error caused by undersampling
is shown for the H-band (F160W) NIC3 PSF (presented in Figure 4).
The stretch now corresponds to 0.12 mag excess to 0.09 mag deficit.}
\label{fig:h_error}
\end{figure}

Figures \ref{fig:v_error} to \ref{fig:h_error} show the error maps
for the WFPC-2 and NIC3 PSFs as
a function of the fractional pixel location of the PSF centroid.
The square area of the maps corresponds to the domain
$-1/2<\delta x<1/2,$ $-1/2<\delta y<1/2,$ in steps of 0.05 pixels;
$(\delta x,\delta y)=(0,0)$ is at the center of the maps.
An important caveat is that in practice finite
limits of integration must be used in equation (\ref{error}),
thus the absolute size of the errors will vary somewhat with aperture.
In the present case I measure the flux in a $1\farcs5\times1\farcs5$
box for the WFC PSFs and a $3\farcs0\times 3\farcs0$ box for NIC3.
Overexposed stellar images show that significant scattered light
falls outside these limits, but in the case of WFPC-2 at least,
the aperture includes nearly all the pixels above the background
for stellar images not saturated in the core.

The WFC error maps show that the photometric effects of undersampling
are subtle, but are still significant for bright sources with sufficient signal.
The error map for the WFC V-band PSF has a peak-to-peak range of 0.030 mag,
and an rms dispersion of 0.008 mag.
The effects of undersampling
are slightly reduced in I-band, as might be expected given its larger PSF width;
the peak-to-peak error range is 0.023 mag, with a 0.006 mag dispersion.
The random color error is thus limited to 0.01 mag.  Intriguingly, however,
the V and I maps qualitatively resemble each other, thus errors in V and
I may correlated depending on how the telescope was pointed for the
two images.  Both filters show that the error
maps are anisotropic in the CCD row ($y$) and column ($x$), with
maximal flux ($\epsilon>0$) actually corresponding to when the PSF
falls between two CCD columns; however, the error maps cannot be simply
described as separable $x$ and $y$ functions.
These results are in excellent agreement
with the simple measurements presented in
\markcite{holtz}Holtzman et al. (1995), who found little dependence
of the photometry on fractional row location, but a few percent variation
dependent on fractional column location, again with more light
detected for stars centered between columns.  Note that this implies
that the intra-pixel response itself for CCDs is more complex than
a simple picture that might have fairly uniform pixels separated by
less sensitive ``cracks.'' \markcite{jdo}Jorden, Deltron \& Oates (1994)
indeed emphasized that at some wavelengths the CCD column stops
corresponded to regions of enhanced sensitivity.
At the same time, they did see this pattern reverse in sign at other
wavelengths --- the reader is cautioned that the present results
are valid only for the F555W and F814W filters.

If undersampling effects are subtle in the WFPC-2 CCDs, they completely
dominate the photometric errors of stellar photometry done with the
NIC3 camera.  The peak-to-peak error range in the NIC3 J-band
is 0.39 mag; the dispersion is 0.10 mag.  The H-band PSF is broader,
given the longer wavelength of the bandpass, thus reducing the
undersampling effects --- still the errors remain large, with
a 0.22 mag peak-to-peak range, and a 0.06 mag dispersion.
The greatest sensitivity indeed occurs for PSFs centered on
a pixel, unlike the case for the WFPC2 CCDs; given the architecture
of the NICMOS arrays, the picture of loosing light in the cracks
between the pixels may be more valid for these devices.

\begin{figure}[htbp]
\plotone{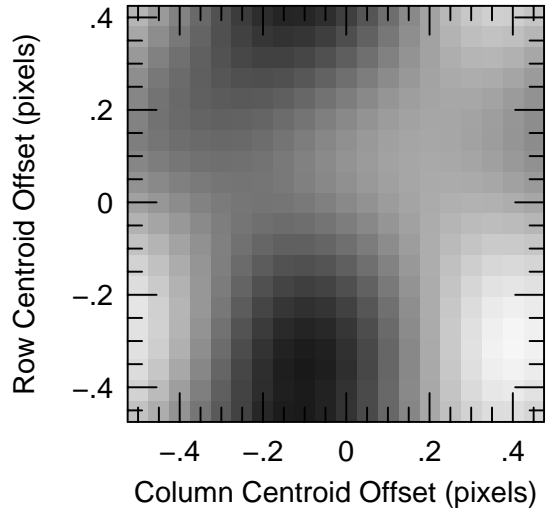}
\caption{The photometric error map for a star also
in the WFPC2 CCD W2 F555W data set, but separated from the star
shown in Figure 1 by 147 pixels.  The stretch is the same as in Figure 5.}
\label{fig:v_error_st2}
\end{figure}

A critical issue is understanding if the present calibrations
can be used to correct for undersampling effects in WFPC2 or NIC3 images.
Unfortunately in the case of WFPC2, it appears that the form of
the PSF core is the dominant contributor to undersampled structure.
Maps made with different stars within the same image set have
error maps that differ significantly from each other.
The WFC PSFs shown above were taken from a bright star near the
center of the W2 CCD of WFPC2.  A map generated from another
star only 147 pixels away in W2 is shown in Figure \ref{fig:v_error_st2}.
Both maps show that the column position dominates the error term,
with maximal flux detected for centroids falling between columns,
and have about same dynamic range.  The detailed structure of the
first map is different enough from the second map, however, such that
it would provide little help for correcting the photometry of a star
imaged at the second location on W2.

\begin{figure}[htbp]
\plotone{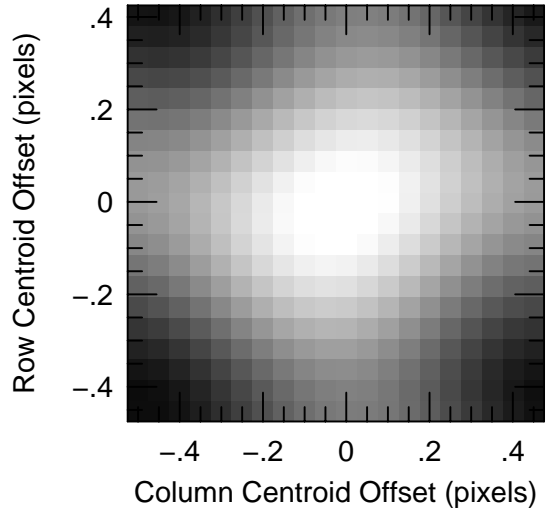}
\caption{The photometric error map for a second star present
in the NIC3 F160W data set, but displaced by about half of the field
from the star shown in Figure 4.
The stretch is the same as in Figure 8.}
\label{fig:h_error_st2}
\end{figure}

The situation is somewhat better in the NIC3 camera, where the pixel
response appears to dominate.
Figure \ref{fig:h_error_st2} shows the map for another star in the
same H-band data set as the star shown in Figure \ref{fig:nic_psf_H}.
The peak-to-peak amplitude of the map is within a few hundredths
of a magnitude of the map shown in Figure \ref{fig:h_error}, and its morphology
is similar enough so that corrections derived from the former star
would work well for the latter.

Regardless of the utility of the present error maps for correcting
undersampled stellar photometry directly, they do show what errors
are likely to be encountered, and how well various dither patterns
sample the error pattern.
The present analysis emphasizes constructing a well-sampled image
from a dither set to counter undersampling present in any single image.
Clearly, a simpler approach of averaging the photometry from a star observed
at different positions in general may reduce the photometric scatter
due to undersampling by $\sqrt{N},$ where $N$ is the number of
dither steps available.
A caveat is that neither the error maps, nor dither pattern are
necessarily random, thus this simple scheme may produce less noise
reduction than expected or still include biases among stars
at differing positions, particularly if the number of dithers is small.
The large range of the NIC3 error maps further imply that an extremely
large data set may be required to obtain 1\%\ photometry by the
simple combination of random dithers.

\begin{figure}[htbp]
\plotone{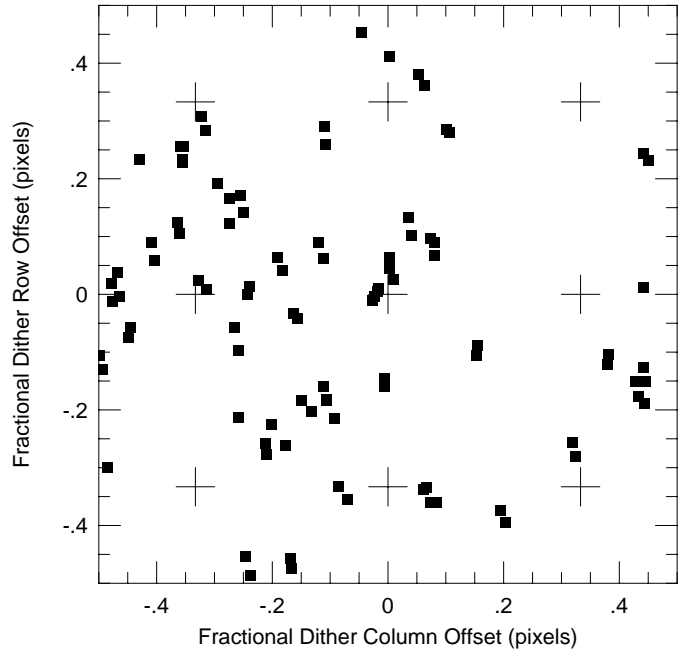}
\caption{The fractional pixel locations of the dithers for the F160W
NIC3 data set are shown.  Crosses mark a putative $3\times3$ dither pattern.}
\label{fig:nic_dithers}
\end{figure}

In this context, I've been surprised by how often dithered
image sets fail to sample the full range of fractional pixel space adequately.
Figure \ref{fig:nic_dithers}, for example, shows the fractional dithers
realized in the NIC3 HDFS image set.
Despite the availability of nearly 100 images, the pattern misses
covering the fractional pixel space as well as a simple regular $3\times3$
pattern would; note that no or very few dithers fall within three
of the corners of the figure.
To be fair, the dithers in the HDFS program were optimized for the
other cameras on board {\it HST}, rather than NIC3, but clearly
the assumption that such a large data set would randomly sample the
full fractional space is not justified.
In the present case, failure to include many dithers landing near the pixel
corners in the case of NIC3 clearly produces a strong bias, since these
are the regions in which the flux deficit due to undersampling is most severe;
further, other stars in the same image set will have their dithers phased
differently, thus biases due to incomplete dither coverage in this
particular image set would be presented as large scatter among
stars at different locations.
Lastly, I note that these biases will not be removed by image
reconstruction algorithms, such as {\it Drizzle,} that simply
redistribute the image flux; in the end one is effectively still
just averaging the stellar images.
The Fourier reconstruction methods that I discuss in \markcite{far}Lauer (1999),
in contrast, can reconstruct an unbiased PSF from even non-optimal
dither patterns, such as that in Figure \ref{fig:nic_dithers}.
The trick is that the complete set of Fourier components that describe
the PSF may still be represented in the dither set and isolated
algebraically, even if it is not optimally encoded in the data.

\subsection{Computing the Centroid of an Undersampled PSF}

While the error maps encode the photometric error
as a function of the fractional location of the PSF centroid, an
important caveat is that measurement of the centroid itself will
be affected by undersampling.  This issue is central
to the concerns of \markcite{king}Anderson \& King (1999), who
discuss methods to obtain high precision relative astrometry on WFPC2
images for the goal of measuring the relative proper motions of
stars within globular clusters.
I will thus not dwell extensively on this issue, myself.
Nevertheless, being able to obtain accurate centroids of PSFs is critical
to constructing the effective PSF in the first place, and then 
evaluating the photometric error of stars at any position in the image set.

\begin{figure}[htbp]
\plotone{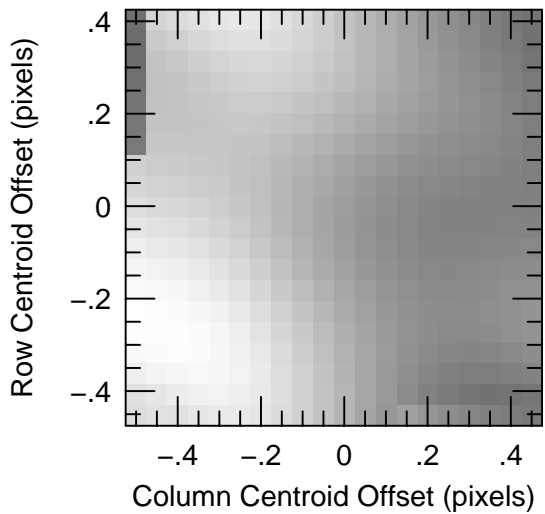}
\caption{An error map showing the total radial difference between
centroids measured from extracted undersampled NIC3 H-band PSFs as a function
of the true fractional displacement of the effective PSF (shown in Figure 4)
from a pixel center.  White corresponds to the maximal error seen of
0.27 pixels.}
\label{fig:h_centroids}
\end{figure}

Figure \ref{fig:h_centroids} gives an error map of the total differences
between the centroids of undersampled NIC3 H-band PSFs (drawn from the
effective PSF presented in Figure \ref{fig:nic_psf_H}) as a function of the true
offset.  In this case, the centroids were computed as the simple
center of weight of a $5\times5$ box centered on the brightest pixel
of the extracted PSF.  The size of the error varied smoothly over the
fractional shift domain, ranging from 0.12 to 0.27 pixels;
if the computed centroids were used to look up the corresponding
H-band photometric error in Figure \ref{fig:h_error}, clearly
the implied photometric correction could be substantially in error.
Now it is true that this simple method of calculating centroids perhaps
would never be the algorithm of choice for undersampled data,
but this is the point --- methods that do work well for
centroiding well sampled PSFs may work poorly for undersampled data,
motivating the use of more sophisticated approaches.

The method that I have used to centroid undersampled PSFs is to
cross-correlate them with a well-sampled PSF.  This, of course, is
fine if one has already generated a PSF from other stars in the image set,
or can fold in knowledge of the pixel response with the construction
of a theoretical PSF (see the next section).  In practice, however,
I've found that the Fourier reconstruction method is fairly
tolerant of centroid errors, and for WFPC2 and NIC3 data, an initial
{\it ad hoc} effective PSF can be constructed from simple centroiding
algorithms as discussed in the previous paragraph.  The penalty
is some blurring in the PSF core, but the {\it ad hoc} effective PSF
can then be used to derive more accurate centroids from the
cross-correlation method --- indeed, this can lead to an iterative
loop where one is continually refining the centroids and the
effective PSF in successive stages.  If a pre-existing effective PSF is available,
however, a critical step is to normalize it as closely as possible to
the expected flux of the new PSF being constructed.
With undersampled PSFs, particularly when much of the flux in contained
in a single bright pixel, positional information is lost
and there can be strong covariance between
intensity scaling and centroid measurement,
a point emphasized by \markcite{king}Anderson \& King (1999).
Again in practice, however, I've found with WFPC2 and NIC3 data that
one can readily construct an initial {\it ad hoc} PSF with a rough
initial normalization.  Lastly, of course, good information on the
dither steps may already be available from external information,
or measurements conducted from an ensemble of other sources in the image set.

\subsection{Isolating the Intra-Pixel Response Function}

Equation (\ref{Ps}) shows that
the effective PSF, $P^\prime,$ is the convolution of the
intrinsic optical PSF with the pixel response, ${\cal R}.$
For the discussion so far, there has been no need to separate the two
terms contributing to $P^\prime;$ however, if knowledge of the optics-only
PSF is available, or it can be calculated by an algorithm such as Tiny Tim
(\markcite{tt}Krist \& Hook 1997), then it may be possible to isolate the pixel
response by deconvolution.  If ${\cal R}$ is largely constant
over the array, as \markcite{jdo}Jorden, Deltron, \& Oates (1994)
suggest is true for CCDs, then is may be possible to use
it in conjunction with theoretical spatially-variable PSFs
to construct improved subsampled PSFs at any point within the field.

\begin{figure}[htbp]
\plotone{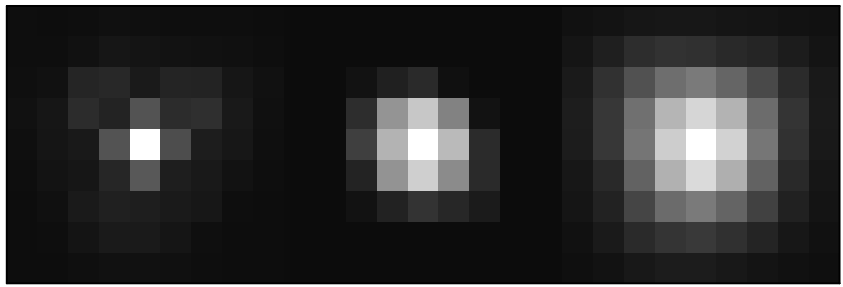}
\caption{The F555W subpixel response for the W2 CCD in shown in the
center with $3\times3$ subsampling.  The total area of each image
is $9\times9$ subpixels, or $3\times3$ full WFC pixels.  The
core of the subsampled reconstructed PSF (the same star as in Figure 1)
is at right, and the Tiny Tim PSF estimate is at the left.  The
stretch is linear, and all three images are normalized to the same
peak intensity.}
\label{fig:subV}
\end{figure}

Figure \ref{fig:subV} shows an attempt to isolate ${\cal R}$
for the F555W filter and the W2 CCD of WFPC2 by
deconvolving the effective PSF in Figure \ref{fig:wfc_psf_V}
with a theoretical PSF constructed with the Tiny Tim package.
The particular star selected has $V-I=1.09$ (in the WFPC2 filters),
which corresponds very closely to spectral type K0.
The theoretical PSF was constructed for the star's location on W2
and with $3\times3$ subsampling.  Deconvolution was done with
160 iterations of Lucy-Richardson deconvolution
(\markcite{rich}Richardson 1972; \markcite{lucy}Lucy 1974);
however, convergence occurred well in advance of this many iterations.

The F555W ${\cal R}$ kernel is in excellent agreement with previous,
but full-pixel rather than sub-pixel estimates of the WFPC2 pixel response.
\markcite{holtz}Holtzman et al. (1995) noted that some diffusion
of light across pixel boundaries appeared to be occurring in the
WFPC2 CCDs.  \markcite{tt}Krist \& Hook (1997) suggest a kernel
that has 75\% of its integral in a central pixel, with 5\% flanking
pixels in the row and column of the central pixel.
The present kernel is given below with $3\times3$ subsampling,
with the pixel values given as percentages of the total integral.
\begin{equation}
\matrix{0.3&1.1&1.7&0.2&0.0\cr
1.8&7.4&10.2&6.4&0.4\cr
2.8&9.1&13.2&9.5&1.7\cr
1.3&7.4&10.6&7.0&1.7\cr
0.4&1.2&2.2&1.5&0.8\cr}
\end{equation}
Pixels outside the $5\times5$ kernel listed are essentially zero;
the diffusion out of the central (full-sized) pixel is
actually limited to only a thin margin a single subpixel in width.

\begin{figure}[htbp]
\plotone{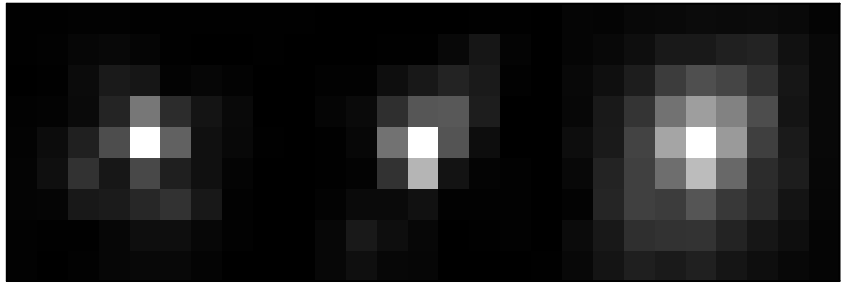}
\caption{The F110W subpixel response for NIC3 in shown in the
center with $3\times3$ subsampling.  The total area of each image
is $9\times9$ subpixels, or $3\times3$ full NIC3 pixels.  The
core of the subsampled reconstructed PSF (the same star as in Figure 3)
is at right, and the Tiny Tim PSF estimate is at the left.  The
stretch is linear, and all three images are normalized to the same
peak intensity.}
\label{fig:subJ}
\end{figure}

Figure \ref{fig:subJ} shows a similar attempt to isolate ${\cal R}$
for the F110W filter in the NIC3 camera.  As expected
given the more severe undersampling effects in NIC3, the
${\cal R}$ kernel is more compact and sharply peaked
than the WF2 pixel response.
The NIC3 pixel kernel is given below with $3\times3$ subsampling,
with the pixel values given as percentages of the total integral.
\begin{equation}
\matrix{0.2&1.2&2.2&3.1&2.3\cr
0.9&4.2&7.5&7.7&2.5\cr
0.6&10.0&22.3&7.4&1.2\cr
0.3&4.4&15.9&1.7&0.4\cr
0.8&0.9&1.5&2.5&0.2\cr}
\end{equation}

\section{Discussion and Summary}

\subsection{Photometry and the Structure of a Pixel}

A one-sentence summary of this paper is that the precision of stellar
photometry may be significantly limited when the PSF is undersampled.
The common assumption that
a CCD consists of an array of contiguous and uniform pixels is
an excellent initial approximation, but is not correct in detail.
One may be tempted to adopt a refined picture in which the array
consists of uniform pixels, but surrounded by dead ``moats;''
however, this still is likely to be an oversimplification.
In truth, the sensitivity pattern within a pixel is likely
to be complex and highly dependent on the specifics of the
detector architecture --- indeed, once one allows for possible
diffusion of photons or photoelectrons within the detector,
the total spatial response function of a single pixel may be
more complex than can be described by pure sensitivity variations alone.
The import of the pixel response depends directly on the severity
of the undersampling and the structural content of the astronomical
source being imaged.  It should also be understood that the pixel
response may be even more important than the core of the optics PSF
in setting the final resolution of an image.

\subsection{Dithering Strategies}

If countering the effects of undersampling on stellar photometry
is important, then I argue that the best solution is to dither
the images in a regular pattern that permits easy reconstruction
of a well-sampled superimage.  The information content of the
superimage is as complete as can be allowed for the particular
properties of the camera's detector and optics.

The optimal dither pattern is a regular $N\times N$ grid of $1/N$
subpixel steps.  In practice one may want to add full integral
steps to the fractional steps as a way of stepping over hot pixels,
bad columns, traps, or any other compact detector defects;
however, if there are significant scale variations over the detectors
field, then it is best to keep the total spatial extent of the
dither pattern as compact as possible.  In an ideal case, one
would also obtain two or more exposures at each dither step,
so as to eliminate cosmic rays events, or any other variable noise feature.
If large angular steps are desirable as well to counter any
large scale variations in the detector response, then I suggest
that the best way to proceed is to obtain the full data set as in subsets
of complete compact dither sequences separated by the larger offsets.
Each subset will make a well-sampled superimage;
combining the superimages into a final image is then simple.

If a regular dither pattern can be executed exactly, then construction
of a superimage requires nothing fancier than simple interlacing
of the individual images.  If the dither positions fall somewhat
away from their optimal locations, but the fractional pixel domain
still has good coverage, or the image set is over-determined,
then I suggest the Fourier method used in this paper as a possible
reconstruction algorithm.  However, even if no formal reconstruction
is attempted, a regular dither pattern will
optimize the information content of the image set.  Lastly,
I emphasize that in general this is a fully two-dimensional problem.
While the WFCP2 pixel response, for example, is more important in
the column direction, it cannot be cleanly separated into separate
$x$ and $y$ functions.  For cameras like NIC3,
two orthogonal one-dimensional patterns will
fall well short of mapping the fractional pixel domain.

\subsection{Designing Undersampled Cameras}

The choice of a pixel scale for an astronomical camera often
requires a compromise between having
a large a field as possible versus obtaining well sampled images.
Since even rather poorly sampled CCD cameras, such as the WFC channel
of WFPC-2, produce excellent stellar photometry for most problems,
it is difficult to argue against tipping their design towards the
largest field that the optics can accommodate.
However, for many of the near and mid-IR cameras contemplated for
space missions now in the early design phases, one must
recognize that IR-arrays may be less forgiving of undersampled PSFs
and may limit the photometric accuracy to unacceptable
levels if the undersampling is too extreme.

If designing a Nyquist-sampled camera causes
unacceptable limitations on the field, however, then I suggest that one
may want to include a dither capability directly in the camera, itself.
{\it HST} has demonstrated the value of dithering undersampled images,
but dithering {\it HST} images is both awkward and prone to
error or non-optimal patterns since the full spacecraft must be moved.
An in-camera dither capability, in contrast, can likely be made to be
simple, highly accurate, and easy to invoke;
ideally, the dither capability would be accurate enough to allow
for direct interlace reconstruction of the superimage.
The caveats are that the dithering must be conducted on timescales
shorter than those on which significant variation in image structure
will occur; when readout noise or overhead becomes significant the minimum
timescale between dithers may make dithering difficult.
Lastly, I emphasize that even fine dithering is unlikely to lift the
effects of severe undersampling.  As a camera becomes increasingly
poorly sampled, not only does the size of the image set required rise
as the square of the inverse dither step (and with it the attendant
difficulty of maintaining a stable image over the length of a dither
sequence), but the precision to which small errors in the pattern
can be detected and corrected for declines, making accurate
image reconstruction all the more difficult.
In practice, pushing beyond $N=3$ subsampling for either
WFPC-2 or NIC3 appears to be highly cumbersome,
and thus may imply final upper limits to the pixel scales
of other cameras in general if accurate photometry is both important and
limited by undersampling.

\acknowledgments

I thank Ken Mighell, Todd Boroson, and Dave Monet for useful conversations.
I thank Jay Anderson and Ivan King for an advance look at their manuscript on
astrometry from undersampled PSFs and many email exchanges, which helped
inform this discussion.
Harry Ferguson kindly provided the reduced NICMOS images used in the analysis.
Alex Storrs kindly provided the Tiny Tim NICMOS PSFs.

\end{document}